\documentclass{article}\begin{document}\centerline{A Note on Lars Onsager and the Partition Functions of Cubic Lattice Models}
\vskip .2in
\centerline{M.L.Glasser}
\centerline{ Department of Physics}
\centerline{\bf Clarkson University, Potsdam NY 13699  USA}
\centerline{\bf laryg@clarkson.edu}
\centerline
{ and}
\centerline{Department of Theoretical Physics}
\centerline{University of Valladolid, Valladolid 40011, Spain}\vskip .5in
\vskip 1in
\centerline{\bf ABSTRACT}\vskip .1in
\begin{quote}
An $n$-dimensional generalization of the Onsager Ising partition function integral is reduced to a single integral and applied to evaluate the partition function and residual entropy of an eight vertex model.\end{quote}
\vskip .4in


\newpage
\noindent
\centerline{\bf Introduction}\vskip .1in

During the decade from 1965-1975 the Nobel Laureate Lars Onsager  served as a technical advisor for Battelle Institute in Columbus, Ohio. At that time I shared an office  there with Dr. Edmund Drauglis, Onsager's ``favorite student" [1] and Lars would drop by several times a year to chat.  After he had seen my gloss [2] on E.H. Lieb's brilliant work [3] on the so called "Ice Models" we found that we shared an interest in  arcane definite integrals, which frequently became the subject of our discussions. Those who knew Lars might remember his penchant for mumbling inaudibly, chortling about what he had said and then scribbling barely decipherable, but relevant, mathematics. On one occasion I suggested that we try to evaluate his famous expression [4] for the free energy
$$-\beta F=\ln 2+\frac{1}{8\pi^2}\int_{[0,2\pi]^2}\; d^2\theta\ln[\cosh^2(2\beta  J)-\cosh(2\beta J)[\cos\theta_1+\cos\theta_2]]$$
 of the 2D Ising model,  which he had left as a double integral. By the end of the discussion we had identified it as a $\;_4F_3$ hypergeometric function, but several key steps were buried in Onsager's mysterious jottings and the result was never published. In 2015  G. Viswanathan [5] derived an equivalent expression obtained by a more comprehensible route.

Recently a paper has appeared  by  De-Zhang  Li,  Yu-Jun Zhao, Yao Yao and Xiao- Bao Yang [6] on an Ising Model including extra interactions, whose partition function is given in Onsager's double integral form. While the aim of this note is to present this  one in  hypergeometric form, an $n$-dimensional version is evaluated by a streamlined version of the unpublished technique and then presented for various values of $n$.\vskip .2in

\centerline{\bf Calculation}

Let 
$$f(a)=\frac{1}{(2\pi)^n}\int_{[0,2\pi]^n}d^n\theta \ln[1-a\sum_{j=1}^n\cos\theta_j]. \quad  Re[a]<1.\eqno(1)$$
Since $f(0)=0$ and, by symmetry,
$$f'(a)=-\frac{n}{(2\pi)^n}.   \int d^n\theta. \frac{\cos\theta_1}{1-a\sum_j\cos\theta_j}$$
$$=-\frac{n}{(2\pi)^n}\int d^n\theta\cos\theta_1\int_0^{\infty}ds e^{-s}\prod_{j=1}^ne^{as\cos\theta_j}=-n\int_0^{\infty} e^{-s} I_1(as) I_0^{n-1}(as)$$
where $I_j$ denotes the modified Bessel function of order $j$. Since $I_0'(x)=I_1(x),$ we have
$$f'(a)=\int_0^{\infty}ds.   \frac{e^{-s}}{s}\frac{d}{da} I_0^n(as)\eqno(2).$$ 
Therefore, by  integrating (2), we recover $f(a)$ as
$$f(a)=\int_0^{\infty} ds \frac{e^{-s}}{s}[I_0^n(as)-1].\eqno(3)$$
and for  the partition function of a "free Fermion hypercubic lattice model" one has
$$Z_n(\alpha,\beta)=\frac{1}{(2\pi)^n}\int_{[0,2\pi]^n}d^n\theta \ln[\alpha-\beta\sum_{j=1}^n\cos\theta_j]=\ln \alpha+f(\beta/\alpha).\eqno(4)$$
The remaining $s$-integral (3)  is 
$$\ln\left(\frac{2}{1+\sqrt{1-a^2}}\right),   \quad a\le1 \quad n=1,$$
$$\frac{1}{2}a^2\;_4F_3(1,1,3/2,3/2;2,2,2;4a^2), \quad a<\frac{1}{2}, \quad n=2,$$
and is related to the cubic [7] and hypercubic [8] lattice Green functions for $n=3,4$. Larger values of $n$ remain to be explored.\vskip .2in
\centerline{\bf Discussion}\vskip .1in

By applying  (4) to equation (13) of reference [4] one obtains 

$$ \lim\; \frac{2}{N}Z_N=\beta(2J-\Delta)+\ln(3+b)-2a\;_4F_3(1,1,3/2,3/2;2,2,2;4a^2)$$
$$a=\frac{(1-b)^2(1+b)}{(3+b)^2},\quad b=e^{8\beta j}.\eqno(5)$$

An elementary expression for  the hypergeometric term in (5)  is not known, but it can be reduced  to elliptic form.  Note that
$$\frac{(1)_n}{(2)_n}z^n=\frac{1}{z}\int_0^z dt t^n$$
so 
$$\;_4F_3(1,1,3/2,3/2;2,2,2;z)=\frac{1}{z}\int_0^z  dt \;_3F_2(1,3/2.3/2;2,2;t),     $$
but  the last integrand is known [9],
$$\;_4F_3(1,3/2,3/2;2,2;z)=\frac{8}{\pi}[{\bf K}(\sqrt{z})-\pi/2]$$

 so 

$$\;_4F_3(1,1,3/2,3/2;2,2,2;z)=\frac{16}{\pi z}\int_0^{\sqrt{z}.}     \frac{{\bf K}(u)-{\bf K}(0)}{u}du$$
$$=\frac{16}{\pi z}\left(\int_0^{\sqrt{z}} \frac{{\bf E}(u)-{\bf E}(0)}{u}du+\frac{\pi}{2}-{\bf E}(\sqrt{z})\right)$$

$$=\frac{16}{\pi z}\left( \int_0^{\frac{\pi }{2}} \log \left(\frac{2}{\sqrt{1-z \sin ^2(t)}+1}\right) \, dt +\frac{\pi}{2}-{\bf E}(\sqrt{z})\right) $$
   $$=  - \left(    \,
   _4F_3\left(\frac{1}{2},1,1,\frac{3}{2};2,2,2;z\right)+\frac{16}{\pi z}E(\sqrt{z})+\frac{8}{z} \right)\, \rm{ if }\,\Im(z)\neq 0\lor \Re(z)<0$$
   For the residual entropy, applying  (4) to [4;(15) ] we get
   $$S/k_B=\ln3-\frac{2}{81}\;_4F_3(1,1,3/2,3/2;2,2,2;\frac{32}{81}).$$
   \vskip .2in
   \noindent
   {\bf Acknowledgement}
   
   This research was supported by Spanish MCIN with funding from European Union NextGenerationEU (PRTRC17.I1) and Consejeria de Educacion from JCyL through QCAYLE project, as well as MCIN project PID2020-113406GB-I00.
\vskip .3in

   \centerline{References}\vskip .1in \noindent
   [1] L. Onsager, Conversation with MLG (1969)
   
   \noindent
   [2] M.L.Glasser,{\it  Evaluation of the Partition Functions for some Two-Dimensional Ferroelectric Models}, Phys. Rev.{\bf 184}, 539-541 (1969)
   
   \noindent
   [3] E.H.Lieb, {\it Exact Solution of the two-dimensional Slater KDP Model of a Ferroelectric},  Phys. Rev. Lett.{\bf 19},  108-110 (1967).
    
   \noindent
   [4]  Lars Onsager, {\it Crystal Statistics I.  A Two-Dimensional Model with an Order-Disorder Transition}, Phys. Rev.{\bf 65}, 117- 123 (1944).
   
   \noindent
   [5] Gandhi  M. Viswanathan, {\it The Hypergeometric Series for the Partition Function of the 2D Ising Model}, J.Statistical Physics: Theory and Experiment, Doi:10.1088/1742.5468/2015/07/P07004
   
   \noindent
   [6]   De-Zhang Li, et al.,{\it  Residual Entropy of a Two-Dimensional Ising Model with Crossing and Four Spin Interaction}, Journal of Mathematical Physics, {\bf 64}, 043303(2023)
   
   \noindent
   [7]  M.L.  Glasser and J. Boersma,{\it  Exact Values for the Cubic Lattice Green Functions}, J. Phys.A:Math. Gen.{\bf 33}, 5017-5023 )2000)\
   
   \noindent
   [8]  M.L. Glasser and A.J. Guttmann,{\it  Lattice Green Function (at 0) for the Hypercubic Lattice}, J.Phys.A:Math.Gen.{\bf  27}, 7011-7012 (1994).
   
   \noindent
   [9]  A.P. Prudnikov et al., {\it Integrals and Series, Vol 3} .[Taylor and Francis,s, NY (2017) Eq. 7.4.2(396).

\end{document}